\documentclass[prb,showpacs,twocolumn]{revtex4}
\usepackage{amsmath,graphicx}

\newcount\minute
\newcount\hour
\newcount\hourMins
\def\now
{
  \minute=\time
  \hour=\time \divide \hour by 60
  \hourMins=\hour \multiply\hourMins by 60
  \advance\minute by -\hourMins
  \zeroPadTwo{\the\hour}:\zeroPadTwo{\the\minute}
}
\def\timestamp
{
  \today\ \now
}
\def\today
{
  \zeroPadTwo{\the\day}-\zeroPadTwo{\the\month}-\the\year%
}
\def\zeroPadTwo#1%
{
  \ifnum #1<10 0\fi
  #1%
}

\date{\timestamp}
\begin{document}

\title{Transport in a Luttinger liquid with dissipation: two impurities}

\author{Leiming Chen}
\author{Zoran Ristivojevic}
\author{Thomas Nattermann}
\affiliation{Institut f\"{u}r Theoretische Physik,
Universit\"{a}t zu K\"{o}ln, Z\"{u}lpicher Str. 77,
50937 K\"{o}ln, Germany}

\begin{abstract}
We consider theoretically the transport in a
one-channel spinless Luttinger liquid
with two strong impurities  in the presence of dissipation.
As a difference with respect to the dissipation free case, where the two impurities fully transmit electrons at resonance points, the dissipation prevents complete transmission in the present situation. A  rich crossover diagram for the conductance  as a function of  applied voltage,  temperature,   dissipation strength, Luttinger liquid parameter $K$ and the deviation from the resonance condition is obtained. For weak dissipation and $1/2<K<1$, the   conduction  shows a non-monotonic increase as a function of temperature or voltage. For strong dissipation the conduction increases monotonically but is exponentially small.

\end{abstract}
\maketitle

\section{Introduction}

Once a quantum mechanical particle traveling in
one-dimensional space hits a potential barrier
formed by an impurity, it is reflected with a
finite probability. This is not longer true in
the case  of two consecutive identical barriers
forming a quantum dot: the  waves reflected at
the first and the second barrier may interfere
destructively such that  there is perfect
transmission. The condition for the latter  reads
$\tan (ka)=-{\hbar^2 k}/({mU_0})$.
Here $m$ and $k$ denotes the mass and the
wavevector of the incoming wave, and $a$ is the
spacing between the two barriers which are here
assumed to be delta functions of strength $U_0$.
For small and
large $U_0$, respectively, this resonance
corresponds to the conditions $ ka=(n+1/2)\pi$
and $ka=(n+1)\pi$, respectively,  where $n\ge0$
is integer. Thus, for large $U_0$ the resonance
happens if the incident particle has the same
energy as one of the bound states formed between
the barriers, provided they are impenetrable.

So far we considered free electrons. In this
paper we will examine the case of interacting
electrons in one dimension where the interaction
can be parametrized by a single  parameter $K$
with $K>1$ $(K<1)$ for repulsive (attractive)
interaction \cite{FiGl96,Gi03}. Then, in the case
of scattering at  an isolated impurity,
attractive (repulsive) interaction  leads to
perfect transmission (reflection) \cite{KaFi92a}
at low energies.

If two consecutive strong impurities are present,
the physics is  influenced by the Coulomb
blockade in the quantum dot. As was shown by Kane
and Fisher \cite{KaFi92b,KaFi92c} and by Furusaki
and Nagaosa \cite{FuNa93}, resonant transmission
of (spinless) electrons is then possible provided
$k_Fa=\left(n+\frac{1}{2}\right)\pi$, where $k_F$
is the Fermi momentum and $n\ge 0$ integer. With
$k_Fa/\pi\equiv q_0$ the background charge
between the two impurities, at resonance the
ground state of the dot is doubly degenerate with
$n_{\pm}=q_0\pm1/2$  particles inside the dot.
Note that this resonance condition is independent
of the impurity strength $U_0$. Tunneling at
resonance is sequential. If one starts with $n_-$
state of the dot, adding a particle to the dot
does not change the Coulomb energy. In a second
tunneling step the particle number in the dot
goes back to $n_-$. In a similar way we may start
with the $n_+$ state and then first decrease
$n_+$ by one, which is followed by a second
electron tunneling into the dot to get back to
the original state $n_+$.

Thus under resonance conditions adding (or
removing) and electron does not change the
Coulomb energy inside the dot.  Renormalization
group analysis of the impurity strength then
shows that perfect transmission is still present
for weak impurities as long as $K>1/4$
\cite{KaFi92b,KaFi92c,FuNa93}. For strong
impurities perfect transmission survives for
$K>1/2$.  The conductance in both cases is given
by $e^2/h$ for spinless electrons. Therefore,
similar to the non-interacting case, the
existence of a second impurities increases the
tendency to perfect transmission for not too
strong repulsive interaction.

Adding an electron  \textit{off} resonance is
accompanied  by an energy increase which has to
be provided either by a thermal bath or by a
finite external voltage. A thermal bath as well
as a finite voltage drop across the dot also
allows sequential tunneling off-resonance, which
leads to power laws of the conductance as a
function of temperature or voltage. If both are
small enough, sequential tunneling is suppressed
and tunneling occurs in one step via the
formation of a virtual state in the dot
(co-tunneling).

In the present paper we want to study the
influence of ohmic dissipation on the scenario
presented so far. As discussed recently in
Ref.~\cite{Cazalilla+06}, ohmic dissipation may
result from the coupling of electrons in the
Luttinger liquid to normal Fermi liquid like
electrons in nearby gates. Under the conditions
considered in \cite{Cazalilla+06} coupling to the
gate is relevant only for $K<K_{\eta}=1/2$.
However, other scenarios are conceivable, and in
the following we will assume that $K_{\eta}$ may
take also larger values. We will therefore assume
that dissipation is present. Clearly, for
$K>K_{\eta}$ our results have to be replaced by
those  of the dissipation free case. of the wire
obeys Dissipation introduces a new length scale
$L_{\eta}\approx 1/(K\eta)$ where $\eta$ denotes
the dissipation strength. On scales larger than
$L_{\eta}$ the plasmon excitations of the
electrons become diffusive and displacement
fluctuations are strongly suppressed, restoring
translational long range order (Wigner crystal).
If $Kv\eta\gg \omega$, the conductivity
$\sigma_{\eta}=2KL_{\eta}e^2/h$ is finite which
is paralleled by diverging superfluid
fluctuations. We have recently shown
\cite{RiNa08} that dissipation has a dramatic
influence on the tunneling of electrons through a
single impurity, which is strongly suppressed.
The voltage and temperature dependence of the
conductance is reduced from power laws in the
dissipation free case to an exponential
dependence for all $K<K_{\eta}$. Thus there is no
region of perfect transmission anymore with a
single impurity. In the present paper we want to
extend these considerations to the case of two
impurities.

The paper is organized as follows. In Sec.~II we
introduce a model for spinless electrons in the
presence of dissipation and with two impurities.
Using the instanton method, we study the electron
tunneling through impurities driven by external
voltage or temperature. The conductance of the
system is calculated in Sec.~III for the
co-tunneling case, and in Sec.~IV for the
sequential tunneling.  Conclusions are drawn in 
Sec.~V. Some technical details are presented in
the appendices.

\section{The Tunneling Probability}

\subsection{The model}
We consider a one-dimensional  interacting system
of spinless electron with two impurities coupled
to a dissipative bath. The impurities  are at
positions $x=\pm a/2$, respectively. We will
refer to the spacing between the impurities as
the \emph{quantum dot}. Using the standard
bosonization methods\cite{Ha81, Gi03}, the
Euclidean action for this system is
\begin{align}\label{Original}
\frac{S}{\hbar}=&\int_{-\frac{L}{2}}^{\frac{L}{2}}dx
\int_{-\frac{\hbar\beta}{2}}^{\frac{\hbar\beta}{2}}d\tau
\bigg\{{1\over 2\pi K}\left[{1\over
v}(\partial_{\tau}\phi)^2+v(\partial_x\phi)^2\right]\\\notag
&+\rho\frac{U_0}{\hbar}\left[\delta(x-a/2)+
\delta(x+a/2)\right]\bigg\}\\\notag &+{\eta\over
4}\int_{-\frac{L}{2}}^{\frac{L}{2}}dx
\int_{-\frac{\hbar\beta}{2}}^{\frac{\hbar\beta}{2}}d\tau
\int_{-\frac{\hbar\beta}{2}}^{\frac{\hbar\beta}{2}}
d\tau'\ \ {[\phi(x,\tau)-\phi(x,\tau')]^2\over
\left[\hbar\beta\sin\frac{\pi(\tau-\tau')}{\hbar\beta}\right]^2}.
\end{align}
Here the displacement field $\phi(x,\tau)$ is
related to the electron density $\rho$ by
\begin{align}\label{rho}
\rho=\frac{k_F}{\pi}-\frac{1}{\pi}\partial_x\phi+\frac{k_F}{\pi}
\cos(2\phi-2k_F x)+\ldots
\end{align}
The first term of action (\ref{Original}) is the
well-known Tomonaga-Luttinger model. The
parameter $K$ measures the interactions between
electrons, where $K<1$ for attractive
interactions and $K>1$ for repulsive
interactions, respectively. $v$ is the velocity
of the plasmon excitations. The second part in
action (\ref{Original}) is the contributions from
the two impurities, where we have assumed that
the two impurities have the same strength $U_0$.

The third piece describes Ohmic dissipation
\cite{CaLe81}. It was shown  in
Ref.\cite{Cazalilla+06} that a dissipation term
of the form
\begin{align}\label{Dissipation}
-\frac{\eta}{2}\int_{-L/2}^{L/2}
dx\int_{-\frac{\hbar\beta}{2}}^{\frac{\hbar\beta}{2}}d\tau
\int_{-\frac{\hbar\beta}{2}}^{\frac{\hbar\beta}{2}}d\tau'
\frac{\cos\left[\phi(x,\tau)-\phi(x,\tau')\right]}{
\left[\hbar\beta\sin\frac{\pi(\tau-\tau')}{\hbar\beta}
\right]^2}
\end{align}
results from the coupling of the electrons in the
wire to Fermi fluid electrons in a nearby gate,
where $\eta$ is a coupling constant. When the
coupling is relevant, Eq.~({\ref{Dissipation})
can be expanded up to quadratic terms, and it has
been done in Eq.~(\ref{Original}).

Integrating out the bulk phase field
$\phi(x,\tau)$ except $\phi(x=-a/2,\tau)$ and
$\phi(x=a/2,\tau)$, we obtain the effective
action as
\begin{align}\label{Impurity}
\frac{S}{\hbar}=&\frac{1}{2K\hbar\beta}\sum_{\omega_n}
\left[\frac{1}{I_+({\omega_n})}|\phi_+(\omega_n)|^2+
\frac{1}{I_-({\omega_n})}|\phi_-(\omega_n)|^2\right]\\\notag
&+U\int
d\tau\cos\left[\phi_+(\tau)\right]\cos\left[\phi_-(\tau)-k_Fa\right]
\end{align}
where
\begin{align}
&\phi_{\pm}(\tau)=\phi(a/2,\tau)\pm\phi(-a/2,\tau),\\
&\phi_{\pm}(\omega_n)=\int d\tau
e^{i\omega_n\tau}\phi_{\pm}(\tau),\quad
\omega_n=\frac{2\pi n}{\hbar\beta},\\
&I_{\pm}({\omega_n})=\frac{\pi\left(1\pm
e^{-{a\over v}\sqrt{\omega_n^2+|\omega_n|v\eta
K}}\right)}{\sqrt{\omega_n^2+ |\omega_n|v\eta
K}}.
\end{align}
$U=2k_F U_0/(\pi\hbar)$ denotes  the
dimensionless pinning strength.


\subsection{Classical Ground State and Excitations}


We will first look at the \textit{classical
ground state}, where $\phi(x,\tau)\equiv\phi(x)$
corresponding to weak quantum fluctuation limit
$K\ll 1$. In fact our further calculation is
strictly justified only in this case although we
will also apply our results for
$K=\mathcal{O}(1)$. The field $\phi_-$ is related
to the charge $Q$  (in units of the elementary
charge) accumulated between the two impurities by
\begin{align}
Q=\int_{-a/2}^{a/2}dx\rho(x)=q_0-\frac{\phi_-}{\pi},
\end{align}
where $q_0=k_Fa/\pi$ denotes the background
charge between the impurities. From
(\ref{Original}), we now obtain
\begin{align}\label{classgroundstate}
&\frac{S_{\mathrm{class.}}}{\hbar}=
\frac{E_c}{2T}(Q-q_0)^2+\frac{2k_F U_0}{\pi
T}\cos\phi_+\cos(\pi Q),
\end{align}
where the Coulomb energy of the quantum dot is
\begin{equation}\label{E-c}
E_c=\frac{1}{\kappa a}\,,
\quad\kappa=\frac{K}{\pi v\hbar}\end{equation}
where $\kappa$ denotes the compressibility. In
the ground state, the action
(\ref{classgroundstate}) has to be minimal. For
the rest of the paper we will assume strong
impurities $U\gg vk_F$, i.e. $U_0/\hbar v\gg 1$,
in agreement with the strong impurity condition
from Eq.~(1). As a result $Q$ has to be an
integer. Then, minimizing the first term in
(\ref{classgroundstate}), one obtains
$Q=Q_0=[q_0]_G$, where $[x]_G$ denotes the
closest integer to $x$. Minimizing the second
term gives $\phi_+=(Q+2m+1)\pi$, where $m$ is an
integer. The corresponding ground state values
for $\phi(\pm a/2)$ are
\begin{align}\label{groundstate-phi}
&\phi(- a/2)=\frac{1}{2}\left[(2m+1)\pi-k_Fa\right], 
\\ \nonumber
&\phi( a/2)=\frac{1}{2}\left[(2m+1)\pi+k_Fa\right] +\pi Q.
\end{align}

In the following it is convenient to express the
background charge in the  dot as
\begin{align}\label{eq:m}
{q_0}={n+{1\over2}}-\Delta,
\end{align}
where $n$ is an integer and $-1/2<\Delta<1/2$.
This gives $Q_0=n$ if $\Delta>0$ and $Q_0=n+1$ if
$\Delta<0$.  For $\Delta=0$  the ground state is
twofold degenerate, allowing $Q_0=n$ and
$Q_0=n+1$, respectively.

As already observed by Kane and Fisher
\cite{KaFi92b} the action
(\ref{classgroundstate}) remains invariant under
the transformation $\phi_+\to \phi_++2\pi$. This
transformation corresponds to the transfer of one
electron from the left to the right LL lead. If
$q_0$ is exactly half-integer, i.e. $\Delta=0$,
there is an additional invariance under the
transformation $\phi_+\to \phi_+ +\pi$ and $Q\to
2q_0-Q$. This transformation corresponds to the
transfer of one electron half-way across the
double barrier structure.  This allows the
sequential tunneling through the quantum dot.

The energy cost for
adding (removing) an electron to (from) the
ground state is given by
\begin{equation}\label{Epm}
E_{\pm}=\frac{E_c}{2}\left[1\pm2\left(Q_0-q_0
\right)\right]
=E_c\left[\pm\Delta+\Theta_H(\mp\Delta)\right],
\end{equation}
where $\Theta_H(x)$ denotes the Heaviside step function.

Let us now consider a sequential tunneling
process in the case $1\gg |\Delta|> 0$. For
$\Delta>0$ the ground state is given by $Q_0=n$.
To transfer an electron through the quantum dot
it first goes from the left lead to the dot,
which requires an energy $E_c\Delta$. In a second
process it tunnels to the right lead which sets
this energy again free. For $\Delta<0$ the ground
state of the dot is $Q_0=n+1$ and the particle
transfer begins with the tunneling of an electron
from the dot to the right lead which costs
$-\Delta E_c>0$ followed by the tunneling of an
electron from the left lead to the dot. Thus, for
sequential tunneling with $\Delta\neq 0$ there is
always one {\it hard} tunneling step which
requires an energy $|\Delta|E_c$. This Coulomb
blockade plays a role  at non-zero temperatures
$T$ or voltages $V_0$ as long as $T,
eV_0<E_C|\Delta|$. Because of the symmetry
between the cases $\Delta>0$ and $\Delta<0$, we
can restrict ourselves in the following to the
case $\Delta>0$.

In addition to the charged excitations there are
also {\it neutral} excitations in the quantum dot
of spacing $\pi \hbar v/a=K/(\kappa a)=KE_c$.
Their maximum energy is   $\hbar \omega_c\approx
\hbar v \Lambda$ where $\Lambda$ is of the order
of the Fermi-momentum $k_F$. If the temperature
$T$  is larger than $KE_c$ the quantization of
the neutral excitations in the quantum dot
becomes irrelevant and the tunneling through the
two impurities will become independent, i.e. the
tunneling is incoherent sequential.
Alternatively, one can say that the coherence
length $L_T=\hbar v/T$ of the displacements of
the electrons in the quantum dot is smaller than
$a$. In the opposite limit the tunneling is
coherent. In the following we will concentrate on
this case.

In our model the neutral excitations are damped.
Plasmons become diffusive on scales larger than
$L_{\eta}=1/(K\eta)$, having a characteristic
life time $L_{\eta}/v$. Phase coherence in the
quantum dot is lost if $L_{\eta}<a$, i.e. if
$KE_c<\Gamma=\hbar v K\eta$ where $\Gamma$
denotes the imaginary part of the plasmon energy.


\subsection{Instanton action}


The tunneling rate $\mathcal{R}$  through the
impurity and hence the current ${I=e{\cal R}}$
can be calculated from the imaginary part of the
free energy
\begin{align}\label{definitionI}
&\mathcal{R}=\frac{2}{\hbar\beta}{\mathrm
{Im}}\ln Z,
\end{align}
where the partition function is
\begin{align}
Z=Z_0+iZ_1=\int
\mathcal{D}\phi_+(\tau)\mathcal{D}\phi_-(\tau)
e^{-S[\phi_+,\phi_-]/\hbar}.
\end{align}
Real part of $Z$ includes the (stable)
fluctuations around the classical ground state.
Since the imaginary part $Z_1$ of the partition
function is small compared with the real part
$Z_0$, the tunneling rate can be written as
\begin{align}
\mathcal{R}\approx\frac{2}{\hbar\beta Z_0}
\textrm{Im}\int \mathcal{D}\phi_+
(\tau)\mathcal{D}\phi_-(\tau)
e^{-S[\phi_+,\phi_-]/\hbar},
\end{align}
where the functional integral is with respect to
the functions $\phi_{\pm}(\tau)$ defined on the
interval $[-\hbar\beta/2,\hbar\beta/2]$ and
satisfying $\phi_{\pm}(-\hbar\beta/2)=
\phi_{\pm}(\hbar\beta/2)$. To calculate $Z_1$ we
follow a method developed by Callan and Coleman
\cite{Callan+}. There exist  saddle point
functions $\tilde\phi_{\pm}(\tau)$ which obey the
equations
\begin{align}
\frac{\delta S[\phi_+,\phi_-]}{\delta\phi_{\pm}
}\bigg|_{\phi=\tilde\phi}=0.
\end{align}
With
$\phi_{\pm}(\tau)=\tilde\phi_{\pm}(\tau)+\delta\phi_{\pm}(\tau)$
the action close to the saddle point  trajectory
can be written  in the form
\begin{widetext}
\begin{align}
&S[\phi_+,\phi_-]-S[\tilde\phi_+,\tilde\phi_-]\approx
\sum_{i,j=\pm}\frac{1}{2}\int d\tau
d\tau'\frac{\delta^2 S}{
\delta\phi_i(\tau)\delta\phi_j(\tau')}\Big|_{\phi=\tilde\phi}
\delta\phi_i(\tau)\delta\phi_j(\tau')=
\sum_{i,j=\pm\atop
{n,m}}V_{i,n;j,m}a_{n,i}a_{m,j}=\sum_n\lambda_nc_n^2,
\end{align}
\end{widetext}
where in the last step we have expanded
$\delta\phi_{\pm}(\tau)$ into a complete set of
orthogonal functions $\psi_n(\tau)$,
$\delta\phi_{\pm}(\tau)=\sum_na_{n,\pm}\psi_n(\tau)$
and then diagonalized the resulting quadratic
form in the $a_{n,\pm}$. One of the eigenvalues,
$\lambda_{n=0}$, has to be negative  to give rise
to the imaginary part. Thus
\begin{equation}
iZ_1=e^{-S[\tilde\phi_+,\tilde\phi_-]/\hbar}{\cal N} \int
\left(\prod_n dc_n\right)
e^{-\sum_n\lambda_nc_n^2/\hbar}.
\end{equation}
One of the eigenvalue has to be  zero
corresponding to a shift of the instanton in the
$\tau$ direction \cite{LaOv84} delivering a
factor $\hbar \beta$. $Z_0$ can be calculated in
the same way with only positive eigenvalues at
the stable saddle point.

Performing this program is very difficult in the
present case. Instead, we will look for an
approximate solution. In this case the saddle
point function   $\tilde\phi(\pm a/2,\tau) $ will
assume their groundstate values
(\ref{groundstate-phi})  everywhere apart from
the  regions where $\tilde\phi(\pm a/2,\tau)$  is
increases by $\pi$. This advancement of $\pi$ is
triggered by the applied external voltage. The
connection between these pieces are narrow kinks
and anti-kinks  of width $\delta\sim  {1/U}\ll
(vk_F)^{-1}$. Thus the saddle point configuration
$\tilde \phi_{\pm}$ is determined in our
approximate scheme by the positions of the kinks
and anti-kinks. A kink-anti-kink pair will be
called an {\it instanton} in the following.

It is sufficient to consider the case when there
is only one instanton at each impurity with
kink--anti-kink spacing equal to $\tau_1$ and
$\tau_2$ on the left and the right impurity,
respectively. To minimize the action we will
assume that the centers of the instantons have
the same value of $\tau$.  With our
parametrization the saddle point
$\tilde\phi_{\pm}$ is now found from the
condition for the instanton action $\partial
S_{\textrm{inst}}(\tau_1,\tau_2)/
\partial\tau_{1,2}\big|_{\tau_{i}=\tau_{i,s}}=0.$
This gives
\begin{align}
I=&e{\cal R}\approx 2e^{-{S_{\rm{inst}}
(\tau_{1,s},\tau_{2,s})}/{\hbar}}\\\notag
&\times{\rm Im}\int' d\vartheta_1d\vartheta_2
\exp^{-\frac{1}{2\hbar}\sum_{i,j=1,2}
\frac{\partial^2S(\tau_1,\tau_2)}
{\partial\tau_i\partial\tau_j}\Big|_{\tau_i=\tau_{i,s}}
\vartheta_i\vartheta_j},
\end{align}
where $\vartheta_i$ denotes deviations from the
saddle point. Here we used $Z_0\approx 1$. The
prime at the integral excludes the integration
over the center of mass of the instanton. Finally
we ignored here a Jacobian factor which describes
the transition from the original field
$\phi_i(\tau)$ to the instanton dimension $\tau$.

Below we will consider only two cases: either
instantons of equal size appear at both
impurities, corresponding to $\tilde\phi_-=0$,
i.e. $\tau_{1,s}=\tau_{2,s}\equiv \tau_s$. This
case will be called {\it co-tunneling}.  Or there
is only one instanton either on the left or the
right impurity, corresponding to
$\tilde\phi_+=\pm\tilde\phi_-$, i.e.
$\tau_{1(2),s}\equiv\tau_s>0$ and
$\tau_{2(1),s}=0$. This case  will be called {\it
sequential} tunneling. Using the  results
obtained previously for  the single impurity
\cite{RiNa08} case, the instanton action at $T=0$
can be written down immediately. It takes the
form
\begin{align}\label{General}
\frac{S_{\mathrm{inst}}}{\hbar}=&\frac{2S_{\mathrm{kink}}}{\hbar}
(\Theta_{H,\delta}(\tau_1)+\Theta_{H,\delta}(\tau_2))
+\frac{2}{K}\left[f(\tau_1)+f(\tau_2)\right]\nonumber\\
&-\frac{eV_0}{2\hbar}(\tau_1+\tau_2)+
|\tau_1-\tau_2|{E_{\mathrm{sign}(\tau_1-\tau_2)}}\frac{1}{\hbar},
\end{align}
where $\Theta_{H,\delta}(x)$ is a step function
of width $\delta$. The different terms have the
following meaning: $S_{\mathrm{kink}}$ denotes
the action of a kink, while
$t=e^{-S_{\mathrm{kink}}/\hbar}$ is the tunneling
transparency of a single impurity. The next term
in (\ref{General}) includes the kink--antikink
interaction with
\begin{align}
f(\tau)=\int_0^{\omega_c} d\omega{\pi\over
I_+(\omega)\omega^2}\left[
1-\cos(\omega\tau)\right].
\end{align}
The following voltage term describes the decrease
of the energy by transferring an electron from
the left to the quantum dot and from there to the
right. Note that the voltage applied to at the
ends of the system is, in general, different from
the voltage at the impurities. However, if the
wire is not too long and the impurities are
strong, both voltages are approximately the same.
Finally, the last term is the contribution from
the Coulomb blockade. Going over to dimensionless
time variable $\tau K eV_0/\hbar=y$ the action
can be rewritten in the form
\begin{align} \label{General2}
&\frac{S(y_1,y_2)}{\hbar}=\frac{2S_{\mathrm{kink}}}{\hbar}
[\Theta_{H,\delta}(y_1)+\Theta_{H,\delta}(y_2)]\\\notag
&+\frac{2}{K}\big[F(y_1)+F(y_2)\big]-\frac{1}{2K}(y_1+y_2)+
\frac{1}{K}|y_1-y_2|\Delta X,
\end{align}
where
\begin{align}\label{Correlation}
F(y)=\int_0^{yZ}\frac{d\Omega}{\Omega}
\frac{\sqrt{1+\frac{yY}{|\Omega|}}\left(1-\cos
\Omega\right)}{1+e^{-\frac{\pi\Omega}{yX}\sqrt{
1+{\frac{yY}{|\Omega|}}}}},
\end{align}
with
\begin{align}\label{Interaction}
X=\frac{E_c}{eV_0},\quad Y=\frac{\Gamma}{KeV_0},\quad Z=\frac{\hbar\omega_c}{KeV_0}.
\end{align}
Here we have used the ratios $X,Y,Z$ of the
relevant energy scales  of the problem. To
calculated the integral we will assume that
always $1, X, Y \ll Z$, i.e. $KeV_0, \Gamma,
KE_c\ll \hbar\omega_c$. The integral can be
approximated by the replacement $1-\cos\Omega
\approx \Theta_H(\Omega-1)$. The calculation is
done in Appendix \ref{InstantonAction} and gives
the final result (\ref{differentcases}).

\subsection{\label{finiteT} Finite Temperatures}

So far we considered the case of zero
temperature. At low but finite temperature the
action and its saddle points are essentially
unchanged, as long as the saddle point for $\tau$
is smaller than $\hbar/T$. For larger $T$ the
tunneling rate is determined from the maximum of
the action taken at $\tau=\hbar/T$
\cite{NaGiDo03}. Again, for sequential tunneling
one of the saddle points of $\tau_{i,s}$
vanishes. For the further discussion it is
convenient, instead of (\ref{Interaction}) to
introduce the following dimensionless parameters
\begin{align}\label{XT}
X_T\equiv {KE_c\over T},\quad Y_T\equiv
{\Gamma\over T},\quad Z_T\equiv
{\hbar\omega_c\over T}.
\end{align}
Accordingly, the dimensionless imaginary time is
redefined as $z=\tau T/\hbar$. This gives,
instead of (\ref{General2}), for the  instanton
action
\begin{align}\label{instanton-action:finiteT}
\frac{S(z_1,z_2)}{\hbar}&=\frac{2S_{\mathrm{kink}}}{\hbar}\left[
\Theta_{H,\delta}(z_1)+\Theta_{H,\delta}(z_2)\right]-{eV_0\over
2T}(z_1+z_2)\nonumber\\
+&\frac{2}{K}\left[F_T(z_1)+F_T(z_2) \right]
+|z_1-z_2|\Delta\frac{X_T}{K},
\end{align}
where $F_T(z)$ is given by
(\ref{differentcasesT}). Note that here either
$z_1=z_2=1$ for co-tunneling or $z_1=1$ and
$z_2=0$ (or vice versa) for sequential tunneling.

\subsection{\label{Cross}Cross-Over between Sequential and Co-Tunneling}


As mentioned already, to find the tunneling rate
and hence the current $I$, we have to calculate
the saddle points $y_{1,s}, y_{2,s}$ of
(\ref{General2}). At $T=0$ the result depends on
$X, Y,Z$ as well as on $\Delta$ and $K$. All
terms in ({\ref{General2}}) are symmetric in
$y_1, y_2$ apart from the last one which
determines the difference between $y_1$ and
$y_2$. These saddle points are calculated in
Appendix \ref{Crossover} in
Eqs.~(\ref{Sequential}) and (\ref{Cotunnel}),
respectively. To find the cross-over line between
sequential tunneling and co-tunneling we have to
equate the saddle point action of the two cases:
\begin{align}\label{CrossOver}
2F(y_s)-\left({1\over 2}-\Delta
X\right)y_s=\frac{2KS_{\mathrm{kink}}}{\hbar}+{4}F(y_c)-y_c.
\end{align}

As it is shown in Appendix \ref{Crossover} the
cross-over between sequential tunneling and
co-tunneling happens at
\begin{align}
X_c\equiv\frac{E_c}{eV_0}
=\frac{1}{2\Delta}\left\{
\begin{array}{ll}
\frac{1}{1+2Y_1},&\frac{\Gamma}{KeV_0}=Y\le1,\\
1-\frac{1}{2+KS_{\mathrm{kink}}/(\hbar\pi
Y)},&\frac{\Gamma}{KeV_0}=Y\gg 1,
\end{array}
\right.
\label{CrossV}
\end{align}
where $Y_1\equiv \left(t^{2K}\over 2\Delta
Z^2\right)^{1\over 1-K}$. The cross-over line is
depicted in Fig.~\ref{VoltageRegime}.

Next we calculate the crossover between
co-tunneling and sequential tunneling for finite
temperatures.
In this case the crossover condition
corresponding to (\ref{CrossOver}) is given by
\begin{align}\label{CrossOverT}
{2\over K}F_T(1)+\Delta\frac{
X_T}{K}=\frac{2S_{\mathrm{kink}}}{\hbar}+{4\over
K}F_T(1).
\end{align}
To solve this equation, we start with the regime
$\Gamma, E_cK\ll T$, in which $F_T(1)$ is given
by the expression case i in formula
(\ref{differentcasesT}). In this regime
Eq.~(\ref{CrossOverT}) leads to
\begin{align}
\Delta X_T=\frac{2KS_{\mathrm{kink}}}{\hbar}+2
\ln{Y_T}\gg 2.
\end{align}
This violates the starting condition $\Gamma,
E_cK\ll T$. Thus the crossover between the
sequential tunneling and co-tunneling is not
possible in this regime. Similarly, it can be
shown that the crossover cannot happen in the
regime $X_T^2/Y_T\ll 1\ll Y_T$. In the remain
three regimes we find self-consistent solutions
for the crossover. These results are summarized
by the following expression
\begin{widetext}
\begin{align}\label{CrossT}
\Delta X_T\approx\left\{
\begin{array}{ll}
 2KS_{\mathrm{kink}}/\hbar,
 &Y_T\ll 1\ll X_T,\\
 2KS_{\mathrm{kink}}/\hbar+\sqrt{2\pi}\left(\sqrt{Y_T}-1\right),
 &1\ll Y_T\ll X_T,\\
 2KS_{\mathrm{kink}}/\hbar+\left[\sqrt{2\pi Y_T}
 +\sqrt{2\pi}\left({Y_T/X_T}-2\right)\right], &1\ll X_T^2/Y_T\ll Y_T.
\end{array}
\right.
\end{align}
\end{widetext}
The various regimes and crossovers between them
for finite temperature are illustrated in Fig.
\ref{TemperatureRegime}.

\section{Co-tunneling}
\subsection{Zero temperature}
In this section we will consider co-tunneling. In
this case the instanton covers both impurities,
$y_1=y_2=y_c$, and the electron will tunnel in
one step through them.
\begin{figure}
\includegraphics[width=0.9\columnwidth]{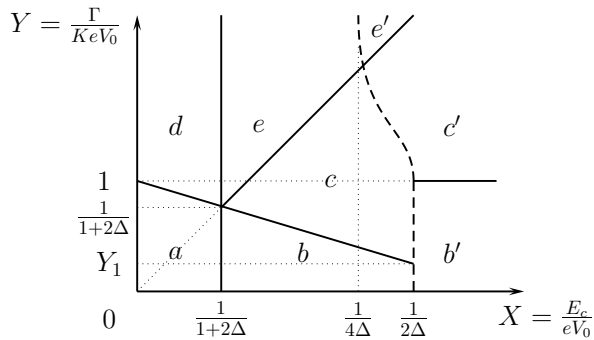}
\caption{Cross-over diagram at zero temperature:
The regions ($a$) - ($e$) correspond to sequential
tunneling, the regions ($b'$), ($c'$) and ($e'$) to
co-tunneling. The dashed cross-over line is given
by Eq.~(\ref{CrossV})}\label{VoltageRegime}
\end{figure}

In the regime of very weak dissipation,
$\Gamma\ll KeV_0$,  i.e. regime ($b'$), we get
\begin{align}\label{R1}
I\sim t^4\omega_{\delta}\left(\frac{KE_c}{\hbar\omega_c}\right)^{\frac{2}{K}}
\left(eV_0\over\hbar\omega_c\right)^{{2\over
K}-1},\quad
{\Gamma}{}\ll KeV_0\ll KE_c
\end{align}
where $\omega_{\delta}\equiv
1/(\omega_c\delta^2)$  and $\delta\sim 1/U$ is a
short time cut-off. This result is similar to the
dissipation free case considered for a single
impurity by Kane and Fisher
\cite{KaFi92a,KaFi92b}. This is intuitively
expected since in the co-tunneling process the
island can be effectively viewed as a ``big''
impurity with renormalized strength. The factor
$t^4$ corresponding to the tunneling through two
impurities. Such a $t^4$ prefactor has been found
previously in a study of Coulomb blockade in a
system with long range interaction \cite{MaGi97}.
For $K>1$ the conductance $G=I/V_0$ diverges
according to (\ref{R1}) which signals the
approach to the perfect conductance $G=e^2/h$.

In the opposite regimes ($c'$) and ($e'$) of strong dissipation, we find
\begin{align}\label{R2}
I\sim  t^4\omega_{\delta}
e^{-\frac{2\pi\Gamma}{K^2eV_0}}
\left(\Gamma\over
eV_0\right)^{3\over 2}&\left(\Gamma\over\hbar\omega_c\right)^{{4\over
K}-1}\left(e^{\frac{\sqrt{2\pi} \Gamma}{E_cK}}-1\right)^{-{2\over K}},\notag\\
&KeV_0\ll{\Gamma}, KE_c.
\end{align}
Again, this result has the same
voltage-dependence as those of the
single-impurity cases\cite{RiNa08}. As it follows
from (\ref{R2}), dissipation strongly reduces the
tunneling probability through the impurities. The
last factor is an interpolation formula between
the cases $\Gamma\ll KE_c$ and $\Gamma\gg KE_c$,
respectively. Since this factor appears as well
in the formulas below,
$t^2\left(e^{\frac{\sqrt{2\pi}
\Gamma}{E_cK}}-1\right)^{-{1\over K}}\approx
t^2e^{-\sqrt{\frac{2}{\pi}}\frac{a}{KL_{\eta}}}$
can be considered as the effective transmission
coefficient of the impurity in the case of strong
dissipation.

\subsection{\label{CotunnelT}Finite temperatures}
\begin{figure}
\includegraphics[width=0.7\columnwidth]{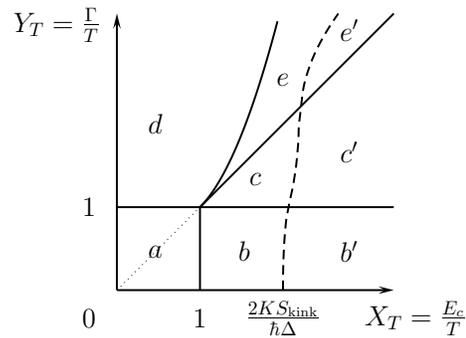}
\caption{Cross-over diagram at nonzero
temperature: The regions ($a$)--($e$) correspond
to sequential tunneling, the regions ($b'$),
($c'$) and ($e'$) to co-tunneling. The dashed
cross-over line is given by
Eq.~(\ref{CrossT}).}\label{TemperatureRegime}
\end{figure}
For nonzero temperature, besides $KE_c$,
$\Gamma$, $\hbar\omega_c$ and $KeV_0$, the temperature $T$ appears as a new
energy scale. This allows in general
a large number of  different regimes. In the following
we will therefore  restrict ourselves to the case
where $KeV_0$ is smaller than all other energies.

In the limit of weak dissipation (regime ($b'$)) we get for the
conductance
\begin{align}\label{R1*}
G\sim t^4\omega_{\delta}
\left(T\over\hbar\omega_c\right)^{{2\over
K}-2},\ \Gamma \ll T\ll KE_c
\end{align}
which again agrees, apart from the factor $t^4$, with the result of Kane and
Fisher \cite{KaFi92a} in the dissipation free case. In the opposite regime ($e'$) of strong
dissipation we get instead
\begin{align}
G\sim  t^4\omega_\delta e^{-\sqrt{\frac{8\pi\Gamma}{K^2T}}}\left(e^{\frac{\sqrt{2\pi} \Gamma}{E_cK}}-1\right)^{-{2\over K}}, T \ll K
E_c,\Gamma
.
\end{align}
It should be noted that the present approach does
not allow the precise determination of the
numerical prefactors in the exponential terms.
The leading temperature dependence is the same as
for the single impurity case, see
Ref.~\cite{RiNa08}.

\section{Sequential Tunneling}
\subsection{Zero temperatures}

Let us next consider the resonant case $\Delta=0$
where the cross-over line between sequential and
co-tunneling moves to $E_c\to\infty$.  Then,
according to (\ref{Epm}), starting with the
ground state $Q=m$, it does not cost energy to
add a particle to the quantum dot.  To remove it
from the quantum dot in the new state $Q=m+1$
does not cost energy as well. The tunneling rate
for each process is the same and follows from the
saddle point of (\ref{General2})  with $y_2=0$
and $y_1=y_s$. Below we will present more general
results for  the case which includes a weak
deviation from perfect resonance.

If  $0<\Delta\ll 1$, bringing an electron to the
quantum dot costs an energy $\Delta E_c$ whereas
in the second step, in which the electron leaves
the dot,  this energy is again released. Thus
tunneling through the dot is dominated by the
first step. Similarly, if $\Delta<0$, it costs
first an energy $|\Delta|E_c$ to bring an
electron out of the quantum dot whereas in the
second step a second electron tunnels from the
left into the dot, which is accompanied by an
energy gain $-|\Delta| E_c$. Thus, again  the
second process is faster than the first one,  the
latter dominates the tunneling probability. Both
cases can be combined by replacing $\Delta$ by
its absolute value.

Plugging the expressions for the saddle points
(\ref{Sequential}) into the action we get
 for the current in the regime ($a$)  of large voltage
\begin{equation}\label{R4}
I=GV_0\sim t^2\omega_{\delta}
\left(eV_0\over\hbar\omega_c\right)^{{2\over
K}-1},\quad {\Gamma}, KE_c\ll KeV_0.
\end{equation}
This corresponds to non-dissipative incoherent
sequential tunneling and has the same voltage
dependence as the single impurity tunneling in
the absence of dissipation \cite{KaFi92a}. The
critical $K$-value for the conductance $G$ is
$K=1$. For $K>1$ the conductance increases for
decreasing voltage signaling a perfect
conductance in the zero voltage limit. It should
however be taken into account that this limit
cannot be performed because of the restriction
$KeV_0\gg \Gamma$.

In the opposite limit of very low voltage (regimes ($c$) and ($e$))
we get instead
\begin{align}\label{R5}
I\sim&
t^2\omega_{\delta}e^{-\frac{\pi\Gamma}{K^2(eV_0-2|\Delta|
E_c)}} \left(\hbar\omega_c\over eV_0-2|\Delta|
E_c\right)^{3\over
2}\\\notag&\times\left(e^{\frac{\sqrt{2\pi}
\Gamma}{E_cK}}-1\right)^{-{1\over K}},
K(eV_0-2|\Delta| E_c) \ll KE_c, {\Gamma}.
\end{align}

In these cases for $|\Delta|=0 $ the  system
shows dissipative resonant tunneling. In
comparison with the corresponding result for the
co-tunneling (\ref{R2}), the expressions in
(\ref{R5}) are larger by an exponent $1/2$ in the
leading voltage dependence (provided $\Delta\to
0$). Clearly, for all values of $K$ dissipation
is dominant and reduces the current strongly.

Finally, there are the intermediate cases ($b$)
\begin{align}\label{R7}
I\sim
t^2\omega_{\delta}\left(\frac{KE_c}{\hbar\omega_c}\right)^{\frac{1}{K}}
\left(eV_0-2|\Delta|
E_c\over\hbar\omega_c\right)^{\frac{1}{K}-1},\\\notag
\frac{\Gamma}{K} \ll eV_0-2|\Delta| E_c \ll E_c
\end{align}
and ($d$)
\begin{align}\label{R8}
I\sim t^2\omega_{\delta}\left(\Gamma\over
\hbar\omega_c\right)^{\frac{2}{K}-1}
\left(\Gamma\over
eV_0\right)^{3\over2}e^{-\frac{4\pi\Gamma}{
K^2eV_0}},\,E_c\ll eV_0 \ll
\frac{\Gamma}{K}.
\end{align}
In case ($b$), Eq.~(\ref{R7}), under resonant
conditions, $|\Delta|=0$, the conductance
$G=I/V_0$ diverges for $V_0\to 0$ and $1/2<K$,
signaling a perfect conductance. For small but
finite $\Gamma$ however, the conductance is
limited by $G\sim
t^2\omega_{\delta}(\Gamma/\hbar\omega_c)^{{1\over
K}-2}$. Case ($d$) corresponds to dissipative
incoherent sequential tunneling. In comparison
with the dissipative single impurity
result\cite{RiNa08}, the expressions in
(\ref{R8}) are smaller by an exponent 2 in the
leading voltage dependence.

Since in regimes $a$ and $d$ the tunneling
through the two impurities is independent, the
total conductance can be calculated by the
formula for two identical conductances connected
in series, $G=G_s(V_0/2)/2$, where $G_s(V_0/2)$
denotes the conductance for single impurity with
voltage drop $V_0/2$. This also leads to results
(\ref{R4}) and (\ref{R8}).

One can notice that in some formulas
(e.g.~(\ref{R2}), (\ref{R8})) no factor $\Delta$
appears, while in (\ref{R5}) and (\ref{R7}) it
appears. The reason is that the latter two
equations correspond to regimes ($b$), ($c$) and
($e$), see Fig.~\ref{VoltageRegime}. In these
regimes, $2\Delta E_c$ can be arbitrary close to
$eV_0$ when one is close to the crossover between
sequential tunneling and co-tunneling. Formula
(\ref{R8}) corresponds to regime ($d$), which is
always far away from the crossover. This implies
that $\Delta E_c$ is always much smaller than
$eV_0$, and hence neglected in (\ref{R8}).
Eq.~(\ref{R2}) is for co-tunneling regime where
the Coulomb blockade term (i.e, the term that
involves $\Delta$ in Eq.~(\ref{General2}))
disappears.

\subsection{\label{SequentialT}Finite Temperature}

At finite temperature we obtain various tunneling
regimes which have one to one correspondence to
those at zero temperature. Moreover, the
discussions for these regimes apply in both cases
of zero and nonzero temperatures. Therefore, to
avoid unnecessary repetition we only give results
for finite temperature. We obtain for the
conductance in regime ($a$)
\begin{align}\label{a}
G\sim t^2
\left(T\over\hbar\omega_c\right)^{{2\over
K}-2}\, ,\quad\Gamma, E_cK\ll T,
\end{align}
which has the same temperature dependence as the
single impurity in the absence of
dissipation\cite{KaFi92a}.

At low temperature we obtain the result of
dissipative resonant tunneling of regimes ($c$)
and ($e$)
\begin{align}\label{Aha}
G\sim t^2 e^{-\frac{E_c|\Delta|}{T}}A_3(T)
e^{-\sqrt{\frac{2\pi\Gamma}{K^2T}}}\left(e^{\frac{\sqrt{2\pi}
\Gamma}{E_cK}}-1\right)^{-{1\over K}},\\ T \ll
KE_c, \frac{(KE_c)^2}{\Gamma} , \Gamma.\nonumber
\end{align}
$A_3(T)$ is some power-law temperature-dependent
function, which is subdominant to the exponential
temperature-dependent part in (\ref{Aha}).

At intermediate temperature we find in regime ($b$)
\begin{align}\label{b}
G\sim t^2
\left(\frac{KE_c}{\hbar\omega_c}\right)^{\frac{1}{K}}
e^{-{E_c|\Delta| \over
T}}\left(T\over\hbar\omega_c\right)^{{1\over
K}-2}, \Gamma\ll T \ll KE_c.
\end{align}
Thus the conductance in regimes ($c$), ($e$) and
($b$) is exponentially suppressed away from
resonance. At resonance (i.e, $|\Delta|=0$) the
conductance in regime ($b$) increases with
decreasing temperature if $K>1/2$, signaling
perfect conductance if $\Gamma=0$. For finite
dissipation the conductance reaches a saturation
value $\sim t^2(\Gamma/\hbar\omega_c)^{{1\over
K}-2}$. Finally in region ($d$) we get
\begin{align}\label{Ahd}
G\sim t^2 A_4(T)e^{-\sqrt{\frac{8\pi \Gamma}{K^2
T}}},\quad \frac{(KE_c)^2}{\Gamma}\ll
T\ll\Gamma
\end{align}
where $A_4(T)$ is again some power-law temperature-dependent function.

\section{Conclusions}
In the present paper we have calculated the
conductance $G$ of a dissipative Luttinger liquid
with a quantum dot formed by two strong
impurities, using an instanton approach. The
following results have been obtained:

(i) Depending on the ratio of the Coulomb energy
of the quantum  dot, $|\Delta|E_c$, and the
temperature $T$ (or the voltage drop $eV_0$,
respectively), there is a crossover from
co-tunneling for low temperatures (or small
applied voltage) to sequential tunneling for
larger temperatures (or voltage). At resonance,
$|\Delta|=0$, the region for co-tunneling
disappears completely and the conductance is
always due to sequential tunneling. The
cross-over lines between co-tunneling and
sequential tunneling are given by
Eqs.~(\ref{CrossV}) and (\ref{CrossT}),
respectively (compare also
Fig.~\ref{VoltageRegime} and
Fig.~\ref{TemperatureRegime}).

(ii) If the voltage drop through the impurities
$eV_0$ is much smaller than all other energy
scales, the response of the system is linear.
Then for very weak dissipation, $\Gamma\ll KE_c$,
and $1/2<K<1$, the conductance at resonance,
$|\Delta|=0$, shows a minimum at $T\approx KE_c$
between the regimes (a) and (b) decribed by
formulas (\ref{a}) and (\ref{b}). This result
agrees with the findings of Furusaki and Nagaosa
\cite{FuNa93}.  For very low temperatures,
$T<\Gamma$,  however, the conductance drops
exponentially due to the dissipation (see
Eq.~(\ref{Aha})  and Fig.~\ref{fig:conductance}.)

Off resonance, for $|\Delta|E_c>T$,  the
conductance is exponentially suppressed even at
larger temperatures, see (\ref{b}). This
reduction of $G$ is limited however by the
cross-over to co-tunneling, see Eq.~(\ref{R1}).

In the opposite limit of strong dissipation,
$\Gamma\gg KE_c$,  the conductance  drops to
exponentially small values as soon as $T\ll
\Gamma$ (see Eqs.~(\ref{Aha}) and (\ref{Ahd})).

(iii) At $T=0$ and at resonance, $\Delta=0$, the
voltage dependent conductance $G=I/V_0$ shows a
behavior similar to that of the temperature
dependent conductance, as follows from eqs.
(\ref{R4}) - (\ref{R8}). There is again a
non-monotonic behavior for $\Gamma<KE_c$ and a
monotonic behavior for $\Gamma>KE_c$ (see also
Fig.~\ref{fig:conductance}).

\begin{figure}
\includegraphics[width=0.7\columnwidth]{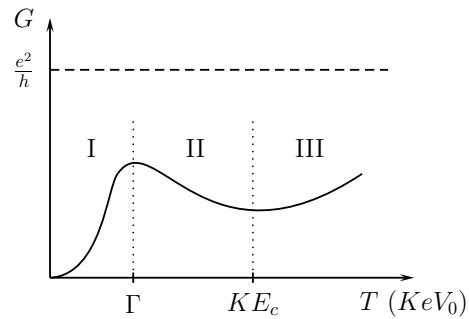}
\caption{Conductance as a function of temperature
(voltage) for the case of weak dissipation
$\Gamma\ll KE_c$.}\label{fig:conductance}
\end{figure}

\subsection*{Acknowledgements}

This work is financially supported by
Sonderforschungsbereich 608. We thank R. Citro
and M. Fogler for discussions in an early stage
of this project.

\appendix

\section{List of symbols}
\begin{tabular*}{\columnwidth}{@{\extracolsep{\fill}} l  l  l  l}
\hline\hline Symbol & Definition & Quantity \\
\hline

$K$ & & Luttinger liquid parameter &  \\

$v$ & & excitation velocity & \\
$\eta$ & & dissipation strength & \\
$\omega_c$ & &high frequency cutoff\\
$a$ & &spacing between impurities\\
$V_0$ & &external voltage& \\
$T$ & &temperature \\
$t$ & &tunneling transparency\\
$\kappa$ &${K}/({\pi\hbar v})$ &
compressibility\\
$E_c$  &${\pi\hbar v}/({Ka})$ &charging energy of the dot\\
$\Delta$ & $n+\frac{1}{2}-\frac{k_Fa}{\pi}$ &distance to the resonance\\
$\Gamma$  &$\hbar v K\eta$& damping of plasmons\\
$X$ & ${E_c}/({ev_0})$\\
$Y$ & ${\Gamma}/({K eV_0})$\\
$Z$ &${\hbar\omega_c}/({KeV_0})$ \\
$X_T$ & ${KE_c}/{T}$\\
$Y_T$ & ${\Gamma}/{T}$\\
$Z_T$ &${\hbar\omega_c}/{T}$ \\

\hline\hline
\end{tabular*}

\section{\label{InstantonAction}}
In this appendix we briefly describe how to
calculate the instanton action for the
interaction between the kink and antikink [i.e.,
$F(y)$ given by formula (\ref{Correlation})]. The
integrant of the integral on the right hand side
of formula (\ref{Correlation}) shows several
crossovers in $\Omega$-space. These crossovers
can be obtained by equating $Yy/\Omega$ and the
exponent of the exponential to 1, separately. For
$Y\ll X$, the crossovers are at $\Omega=Yy$ and
$\Omega=Xy$, which separates the three regions
$\Omega\ll Yy\ll Xy$, $Yy\ll\Omega\ll Xy$, and
$Yy\ll Xy\ll \Omega$. For $Y\gg X$, the
crossovers are at $\Omega=Yy$ and
$\Omega=X^2y/Y$, which again separates three
regions $\Omega\ll X^2y/Y\ll Yy$,
$X^2y/Y\ll\Omega\ll Yy$ and $X^2y/Y\ll Yy\ll
\Omega$. To evaluate the integral we dissemble it
into small ones according to the different
regions separated by the crossovers. We consider
the two cases $Y\ll X$ and $Y\gg X$, separately.
We start with $Y\ll X$. For $1\gg Xy\gg Yy$, we
have
\begin{align}
F(y)\approx \int_0^{Zy} d\Omega
{1-\cos\Omega\over\Omega}=\ln\left(Zy\right).
\label{Small}
\end{align}
For $Yy\ll 1\ll Xy$, we have
\begin{align}
F(y)&\approx {1\over 2}\int_0^{Xy} d\Omega
\frac{1-\cos\Omega}{\Omega}+\int_{Xy}^{Zy}
d\Omega{1-\cos{\Omega}\over\Omega}\notag\\&={1\over
2}\ln\left(Xy\right) +\ln\left(Z\over X\right).
\end{align}
For $1\ll y Y\ll y X$, we have
\begin{align}
F(y)&\approx{1\over 2}\int_0^{Yy} d\Omega
{\sqrt{Yy\Omega}\over\Omega^2}\left(1-\cos\Omega\right)\\\notag
&+{1\over 2}\int_{Yy}^{Xy} d\Omega
{1-\cos{\Omega}\over\Omega}+\int_{Xy}^{Zy}
d\Omega{1-\cos{\Omega}\over\Omega}\\\notag
&={1\over 2}\left[\sqrt{2\pi}
\left(\sqrt{Yy}-1\right)+\ln\left(X\over
Y\right)+2\ln\left(Z\over X\right)\right].
\end{align}

Now we discuss the case $Y\gg X$. For $1\gg Yy\gg X^2y/Y$, one also gets
the result formula (\ref{Small}). For $X^2y/Y\ll 1\ll Yy$, one gets
\begin{align}
F(y)&\approx \int_0^{Yy}
d\Omega{\sqrt{Yy\Omega}\over\Omega^2}\left(1-\cos{\Omega}\right)+\int_{Yy}^{Zy}
d\Omega \frac{1-\cos{\Omega}}{\Omega}\notag\\
&=\sqrt{2\pi}\left(\sqrt{Yy}-1\right)+\ln\left(Z\over
Y\right).
\end{align}
For $1\ll X^2y/Y \ll Yy$, one finds
\begin{align}
F(y)&\approx{1\over 2}\int_0^{X^2y/Y}
d\Omega{\sqrt{Yy\Omega}\over\Omega^2}(1-\cos{\Omega})\notag\\
+&\int_{X^2y/Y}^{Yy} d\Omega
{\sqrt{Yy\Omega}\over\Omega^2}(1-\cos{\Omega})+\int_{Yy}^{Zy}
d\Omega\frac{1-\cos{\Omega}}{\Omega}\notag\\
&={1\over 2}\left[\sqrt{2\pi
Yy}+\sqrt{2\pi}\left({Y\over
X}-2\right)+2\ln\left(Z\over Y\right)\right].
\end{align}
Finally, the results can be summarized by
\begin{widetext}
\begin{align}
 F(y)=\left\{
  \begin{array}{ll}
   \ln\left(Zy\right), &\mathrm{(i)}\quad yY, y X\ll 1,\\
   {1\over 2}\ln\left(Z^2y/X\right), &\mathrm{(ii)} \quad yY \ll 1\ll yX,\\
   {1\over 2}\left[\sqrt{2\pi Yy}+\ln\left(Z^2\over X Y\right)\right], &\mathrm{(iii)}\quad 1\ll yY\ll yX,\\
   \sqrt{2\pi Yy}+\ln\left(Z\over Y\right), &\mathrm{(iv)} \quad  yX^2/Y\ll 1\ll yY,\\
   {1\over 2}\left[\sqrt{2\pi Yy}+\sqrt{2\pi}{Y\over X}+2\ln\left(Z\over Y\right)\right], &\mathrm{(v)} \quad 1\ll yX^2/Y\ll yY.
  \end{array}
 \right.
 \label{differentcases}
\end{align}
\end{widetext}
\section{\label{Crossover}}

The saddle points for sequential and co-tunneling are given by (\ref{Sequential}) and (\ref{Cotunnel}), respectively.
\begin{align}
 y_s=\left\{
  \begin{array}{ll}
   {4\over 1-2\Delta X}, &\mathrm{(a)}\quad Y+2\Delta X, X(1+2\Delta )\ll 1,\\
   {2\over 1-2\Delta X}, &\mathrm{(b)} \quad Y+2\Delta X \ll 1\ll X(1+2\Delta),\\
   {2\pi Y\over (1-2\Delta X)^2}, &\mathrm{(c)}\quad 1\ll Y+2\Delta X\ll X(1+2\Delta),\\
   {8\pi Y\over (1-2\Delta X)^2}, &\mathrm{(d)} \quad  X(1+2\Delta)\ll 1\ll Y+2\Delta X,\\
   {2\pi Y\over (1-2\Delta X)^2}, &\mathrm{(e)} \quad 1\ll X(1+2\Delta)\ll Y+2\Delta X.
  \end{array}
 \right.
 \label{Sequential}
\end{align}
Here the different areas of validity ($a$)--($e$)
are separated by the lines $X=Y,\,\,Y=1-2\Delta
X\,\, \mathrm{and }\,\, X=1/(1+2\Delta)$ (see
also Fig.~\ref{VoltageRegime}). These saddle
points have to be compared with the saddle points
for co-tunneling
\begin{align}
 y_c=\left\{
  \begin{array}{ll}
   4, &\mathrm{(a')}\quad Y, X\ll 1,\\
   2, &\mathrm{(b')} \quad Y \ll 1\ll X,\\
   2\pi Y, &\mathrm{(c')}\quad 1\ll Y\ll X,\\
   8\pi Y, &\mathrm{(d')} \quad  X\ll 1\ll Y,\\
   2\pi Y, &\mathrm{(e')} \quad 1\ll X\ll Y,
  \end{array}
 \right.
 \label{Cotunnel}
\end{align}
with the areas of validity separated by the lines
$X=Y,\,\,Y=1\,\, \mathrm{and }\,\, X=1$ (see
again Fig.~\ref{VoltageRegime}). Clearly, for
$\Delta=0$ both sets of saddle points are
identical, but the action of the co-tunneling
process is always larger than that of sequential
tunneling and hence sequential tunneling
prevails. Qualitatively, this remains true for
small but finite $\Delta X\ll 1$. However, if
$\Delta X$ becomes of the order one, the saddle
points for
sequential tunneling move to larger values such that 
for 
$X>X_c(Y)$ co-tunneling sets in.

The crossover between the sequential tunneling
and the co-tunneling is defined as the point at
which the currents for the two different
tunneling mechanism are equal. Assuming
$\Delta\ll 1$, in Sec. \ref{Cross} it is pointed
out that the crossover between the co-tunneling
and the sequential tunneling in regimes $a$ and
$d$ [i.e., $X\gg 1/(1+2\Delta)$] cannot happen.
This is clearly illustrated in
Fig.~\ref{VoltageRegime}.

We start with very small $Y$ (i.e., $Y<Y_1$) such
that near the crossover the currents for the
sequential tunneling and the co-tunneling are
given by formulae (\ref{R7}) for regime $b$ and
(\ref{R1*}) for regime $b'$, respectively. The
value of $Y_1$ will be determine afterwards.
After a straightforward calculation the crossover
is found to be
\begin{align}
X={1-\left(t^{2K}\over 2\Delta Z^2\right)^
{1\over 1-K}\over 2\Delta}\approx {1\over
{2\Delta}}. \label{22}
\end{align}
However, this result is self-consistent only if
the current for the sequential tunneling is
indeed given by formula (\ref{R7}) for regime $b$
[i.e., $Y+2\Delta X\ll 1\ll X(1+2\Delta)$]. This
leads to a restriction on the validity of
(\ref{22})
\begin{align}
 Y\ll Y_1\equiv \left(t^{2K}\over 2\Delta Z^2\right)^{1\over 1-K}.
\end{align}
$Y_1$ is essentially very small.

For $Y$ slightly bigger than $Y_1$, the current
for the sequential tunneling is given by formula
(\ref{R5}) for regime $c$, while the current for
the co-tunneling remains in regime $b'$. In this
situation the crossover is determined by
\begin{align}
&t^2\omega_{\delta}\left(Y\over Z\right)^{{2\over
K}+{1\over 2}}\left(X\over Y\right)^{1\over
K}Z^{3\over 2}\left(1-2\Delta X\right)^{-{3\over
2}}\\\notag&\times\exp\left(-{\pi Y\over
K(1-2\Delta X)}\right)=
{t^4\omega_{\delta}}\left(X\over Z\right)^{2\over
K}\left(1\over Z\right)^{{2\over K}-1},
\end{align}
where we have chosen to express the currents in
terms of the ratios of relevant energy scales
$X$, $Y$ and $Z$. After some algebra we obtain
\begin{align}
X\approx {1-{\hbar\pi Y\over
2KS_{\mathrm{kink}}}\over 2\Delta}\approx {1\over
2\Delta}.
\end{align}
This result is valid for $Y_1\ll Y\ll 1$.

Now for $1\ll Y\ll X$, the current for the
sequential tunneling is still in regime $c$,
while the one for the co-tunneling just moves
into regime $c'$ and is given by formula
(\ref{R2}). In this regime the crossover is given
by
\begin{align}
t^2\omega_{\delta}\left(Y\over Z\right)^{{2\over
K}+{1\over 2}}\left(X\over Y\right)^{1\over
K}Z^{3\over 2}\left(1-2\Delta X\right)^{-{3\over
2}}\notag\\\times\exp\left(-{\pi Y\over
K(1-2\Delta X)}\right)\notag\\
=t^4\omega_{\delta}\left(Y\over Z\right)^{{4\over
K}+{1\over 2}}\left(X\over Y\right)^{2\over
K}Z^{3\over 2}\exp\left(-{2\pi Y\over K}\right),
\end{align}
which leads to
\begin{align}
1-2\Delta X&=\frac{\pi
Y}{\frac{2KS_{\mathrm{kink}}}{\hbar}+2\pi
Y+\ln\left(Z^3\over XY^2\right)}\notag\\
&\approx \frac{\pi
Y}{\frac{2KS_{\mathrm{kink}}}{\hbar}+2\pi Y}.
\end{align}
This crossover implies that for $1\ll Y\ll X$ the
range of $X$ on the crossover is within
$1/(4\Delta)\ll X\ll 1/(2\Delta)$. Thus, for
$KS_{\mathrm{kink}}/\hbar\ll 1/\Delta$ the
crossover intersects with the line $Y=X$ at
$Y=X\approx 1/(4\Delta)$; otherwise they meet at
$Y=X\approx 1/(2\Delta)$.

For $Y\gg X$, the current for the sequential
tunneling and the co-tunneling are given by
formulae (\ref{R5}) for regime $e$ and (\ref{R2})
for regime $e'$. The crossover is given by
\begin{align}
&t^2\omega_{\delta}\left(Y\over Z\right)^{{2\over
K}+{1\over 2}} Z^{3\over 2}\left(1-2\Delta
X\right)^{-{3\over 2}}\\\notag
&\times\exp\left(-{\pi Y\over K(1-2\Delta
X)}\right)\exp\left(-{\sqrt{2\pi}Y\over
KX}\right)\notag\\\notag
=&t^4\omega_{\delta}\left(Y\over
Z\right)^{{4\over K}+{1\over 2}}Z^{3\over
2}\exp\left(-{2\pi Y\over
K}\right)\exp\left(-{2\sqrt{2\pi}Y\over
KX}\right)
\end{align}
This leads to
\begin{align}
1-2\Delta X&={\pi Y\over
\frac{2KS_{\mathrm{kink}}}{\hbar}+2\pi
Y+\sqrt{2\pi}{Y\over
X}+2\ln\left(Z\over Y\right)}\notag\\
&\approx{\pi Y\over
\frac{2KS_{\mathrm{kink}}}{\hbar}+2\pi Y}.
\end{align}
Finally, the crossover between the sequential
tunneling and the co-tunneling can be summarized
as formula (\ref{CrossV}). The various regimes
and the crossovers between them are illustrated
in XY plane in Fig.~\ref{VoltageRegime}.


\section{\label{Temp}}

In this appendix we quote the final result for
the function $F_T(z)$, which first appears in
Eq.~(\ref{instanton-action:finiteT}). It reads
\begin{widetext}
\begin{align}
F_T(z)=\left\{
\begin{array}{ll}
\ln\left(Z_Tz\right), &\mathrm{(i)}\quad zY_T, z X_T\ll 1,\\
{1\over 2}\ln\left(Z_T^2z/X_T\right), &\mathrm{(ii)} \quad zY_T \ll 1\ll zX_T,\\
{1\over 2}\left[\sqrt{2\pi Y_Tz}+\ln\left(Z_T^2\over X_T Y_T\right)\right], &\mathrm{(iii)}\quad 1\ll zY_T\ll zX_T,\\
\sqrt{2\pi Y_Tz}+\ln\left(Z_T\over Y_T\right), &\mathrm{(iv)} \quad  zX_T^2/Y_T\ll 1\ll zY_T,\\
{1\over 2}\left[\sqrt{2\pi
Y_Tz}+\sqrt{2\pi}{Y_T\over
X_T}+2\ln\left(Z_T\over Y_T\right)\right],
&\mathrm{(v)} \quad 1\ll zX_T^2/Y_T\ll zY_T.
\end{array}
\right.\label{differentcasesT}
\end{align}
\end{widetext}

In the case of finite temperature, the previous
saddle point solution for zero temperature and
finite voltage does not apply when the distance
between the kink and antikink is larger than the
size of the imaginary time axis. Thus the maximum
instanton action occurs at $\tau=\hbar/T$, i.e.,
$z=1$. Then the tunneling rate can be
approximated as proportional to
$\exp\left[-S(z_1=1, z_2=0)/\hbar\right]$ for the
sequential tunneling and
$\exp\left[-S(z_1=1,z_2=1)/\hbar\right]$ for the
co-tunneling, respectively. Unlike in the case of
zero temperature, the tunneling rates along both
the voltage-favored and -unfavored directions are
comparable. Therefore, the current should be
proportional to the difference between these two.
To the lowest order of $eV_0$, we get
\begin{align}
 I \sim eV_0 t^2\exp\left[-{2\over K}F_T(1)-\frac{\Delta X_T}{K}\right]
\end{align}
for the sequential tunneling and
\begin{align}
 I \sim eV_0t^4\exp\left[-{4\over K}F_T(1)\right]
\end{align}
for the co-tunneling. This approximation is not
accurate enough to give correct power-law
temperature-dependence of the current. However,
in the dissipative regime it captures the
dominant exponential temperature-dependence.
Plugging the expressions of $F_T(1)$ into the
above two formulae, we obtain the results given
in Secs. \ref{SequentialT} and \ref{CotunnelT}.

\end{document}